# The driving force governing room temperature grain coarsening in thin gold films

O. Glushko[1*], M. J. Cordill[1,2]

1) Erich Schmid Institute of Materials Science, Austrian Academy of Sciences, Jahnstrasse 12, A-8700 Leoben, Austria

2) Department of Materials Physics, Montanuniversität Leoben, Jahnstrasse 12, A-8700 Leoben, Austria

**Abstract** Strong room-temperature grain coarsening in gold films on polyimide induced by cyclic uniaxial mechanical strain is demonstrated. Detailed electron backscatter diffraction analysis revealed that, in contrast to the predictions of shear-coupled grain boundary migration model, the grain coarsening is isotropic and coarsened grains do not exhibit any specific crystallographic orientations or misorientations to the neighboring grains. It is shown that a thermodynamic model where the driving force appears due to the difference in yield stresses between the grains with different sizes provides an adequate explanation of the experimental data.

Room temperature grain coarsening induced by mechanical loading has been shown to occur in Cu [1-3], Pt [4], Ni [5], Al [6-8], and Au [9] under different loading conditions such as nanoindentation [1,6], monotonic tensile loading [7,8], fatigue loading [2-4,9] as well as beam bending [5]. Despite the large number of experimental evidences, very little is known about the driving forces behind the athermal grain coarsening. In systematic investigations of nanocrystalline Al thin films [7,8] the authors were not able to explain the observed grain coarsening with "traditional driving forces"[7] and it was suggested that shear coupled grain boundary migration (SCGBM) is responsible for the grain growth. However, the SCGBM concept was developed to explain the grain boundary (GB) migration at elevated temperatures in bicrystals with clearly defined GB planes, GB types and misorientations [10-12]. In polycrystals, the grain boundaries are often of mixed type and have no coincidence site lattices. Although the SCGBM was generally observed in polycrystals [13,14], the tangential displacement of the grains with respect to the boundary must be restricted due to the constraint caused by the neighboring grains [15]. Thus, it is currently unclear to which extent the SCGBM concept can be applied to explain the grain coarsening effect in real polycrystals. It is necessary to note, that grain coarsening can also occur without GB migration as demonstrated for nanocrystalline Au films [9] where a nanotwin-assisted GB elimination mechanism was proposed.

In the present work, a detailed electron backscatter diffraction (EBSD) characterization is utilized to analyze possible driving forces governing severe room temperature grain coarsening in cyclically loaded ultra-fine grained (UFG) gold films.

The gold films were deposited on 50 μm polyimide Upilex substrates by electron beam evaporation in a Balzers BAK 550 evaporation machine with the vacuum of $2.1 \times 10^{-7}$ mbar and using a deposition rate of 0.3 nm/s to a thickness of 500 nm. The samples with the width of 5 mm and length of 40 mm were cut out of the sheet using a scalpel. Cyclic tensile straining was performed on an MTS Tytron 250 tensile testing device. Sine strain function between zero strain and a peak strain was applied with the frequency of 0.1 Hz. Such a low straining rate was used in order to account for viscoelastic relaxation of the substrate and to exclude possible heating effects. Three different peak strain values were used in mechanical tests (1%, 1.5%, and 2%) but for the sake of brevity only the results for 1.5% peak strain are shown here. The strain and displacement rates for 1.5% peak strain were 0.003 s$^{-1}$ and 60 µm/s, respectively. After 1000, 2000, and 5000 cycles, the mechanical test was interrupted to perform EBSD analysis and scanning electron microscopy of the surface. Focused ion beam (FIB) marker was used to locate the same area within the film. The points of the EBSD scans which were not indexed properly and have the confidence index of less than 0.05 were removed and appear in black in Fig. 1 and Fig. 2.

An overview of the severe room temperature grain coarsening effect in 500 nm thick gold films on polyimide substrate is shown in Fig. 1. In Fig. 1a the grain orientation map of the initial microstructure in the direction normal to the surface (ND) is shown. The film has a strong (111) texture and average grain size of 210±60 nm. Fig. 1b shows the same surface area after applying 5000 cycles with 1.5% tensile strain. Strong grain coarsening leads to an increase of average grain size, extracted from the same area, to 1420±320 nm. The total number of grains decreased from 9400 to 1400 and there are virtually no surface areas which conserved the initial microstructure. The orientation distributions within the stereographic triangles shown in the insets demonstrate that no significant texture transition occurs during grain coarsening.

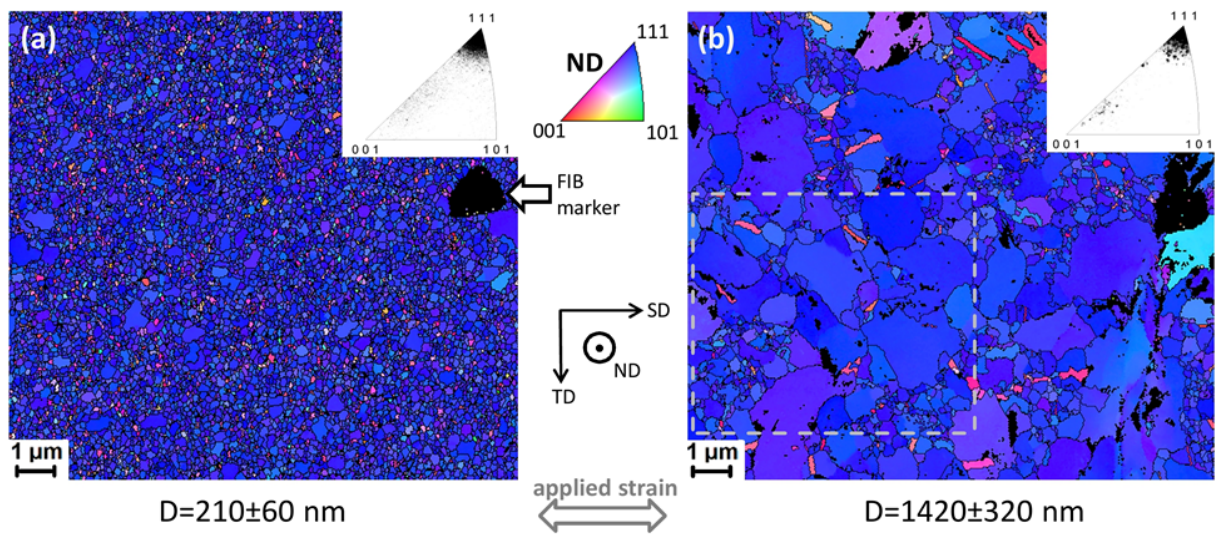

**Fig. 1.** Grain orientation maps of 500 nm thick gold film (a) before and (b) after application of 5000 cycles with 1.5% strain. The insets display the corresponding distributions of crystallographic orientations in stereographic triangles. The arrows marked with acronyms SD, ND, and TD show the strain direction, normal direction, and transverse direction, respectively. The dashed rectangle in (b) corresponds to the area which is considered in detail in Fig. 2.

The most straightforward way to uncover the mechanism of the observed grain coarsening is to determine the common features the coarsened grains have in comparison to non-coarsened grains. To consider the evolution of the microstructure in more detail the grain orientation maps in the straining direction before cyclic loading, after 1000 cycles, and after 5000 cycles are shown in Fig. 2. The grain boundaries are separated into three groups according the misorientation angle. The low angle grain boundaries (LAGB) appear in yellow, the general high angle grain boundaries (HAGB) are shown in black and the twin boundaries are highlighted by the red color. The numbers from 1 to 10 depict the same grains at each stage of the mechanical loading. The first observation which should be made is that the coarsened grains do not have any specific crystallographic orientation with respect to the loading direction. Secondly, the grain coarsening is isotropic, meaning that there is no preferential direction of the grain extension although the applied strain is uniaxial. Third, there is no clear correlation between the grain coarsening and GB misorientation angle. By comparing Figs. 2b and 2c one can find numerous examples of non-migrating LAGBs (e.g. GB surrounding grain 6), HAGBs (e.g. GB between grains 7 and 9) and twin boundaries (e.g. bottom-right GB of grain 5). At the same time, there are many examples of migrating LAGB, HAGB, and twin boundaries. Moreover, the misorientation of the GB of a growing grain changes during coarsening as soon as a neighboring shrinking grain disappears. Thus, the GB misorientation cannot be a decisive factor which promotes the grain coarsening. The only common feature of the coarsened grains, which was determined using the current experimental method, is the initial grain size. Almost all numbered grains have initial sizes, which are more than two times higher than the average grain size. The only exception is grain 9, which has an initial equivalent diameter of 350 nm.

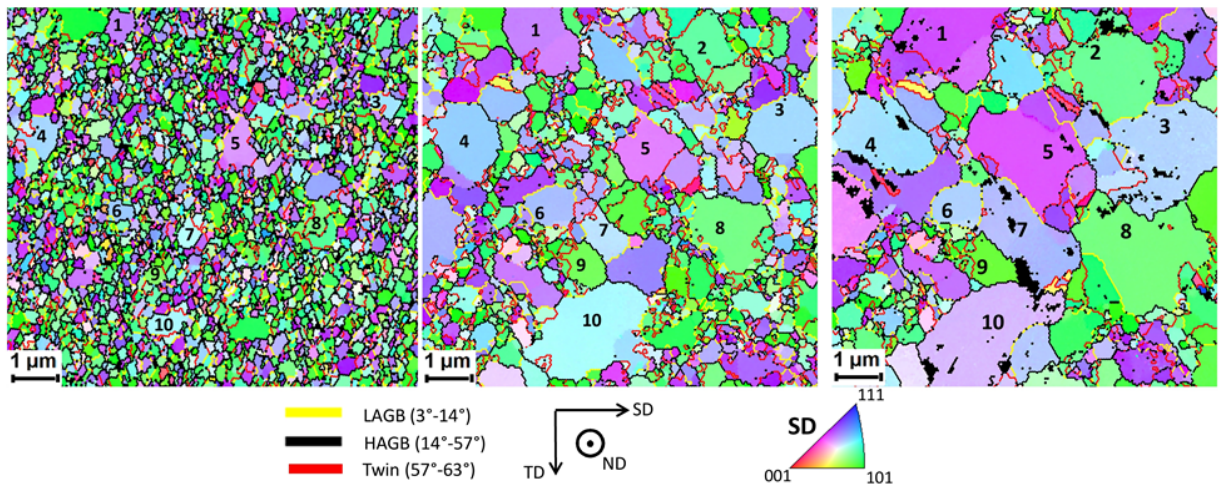

**Fig. 2.** Detailed evolution of the microstructure during cyclic loading. The grain orientation maps in the strain direction (SD) are shown for 500 nm thick gold film (a) before straining, (b) after 1000 cycles (c) and after 5000 cycles with 1.5% peak strain. Selected grains which coarsen during the deformation are marked with the numbers 1 to 10. The low-angle, high-angle and twin grain boundaries are coded by different colors as stated.

Summarizing the features of the grain coarsening effect, one can state that initially larger grains coarsen preferably, independent from the grain orientation or GB misorientation. As a consequence of this statement, the grain coarsening observed here cannot be explained in the frameworks of SCGBM approach due to the following reasons. SCGBM predicts a strong dependence of the normal GB displacement on the GB type, symmetry and misorientation angle. It was even shown that there is a threshold misorientation angle of 36.9° where GB migration changes its direction [10] which is in contrast to the case presented (Fig. 2). SGGBM assumes that GB migration is driven by shear stress which would lead to anisotropic grain coarsening under uniaxial loading conditions. There are also no evidences of the tangential displacement of the grains behind the migrating boundaries predicted by SCGBM. Finally, the fact that the initial grain size promotes grain coarsening is also not supported by SCGBM where the grain size is not considered as a parameter.

The grain coarsening effect presented here can be explained in the frameworks of a more general thermodynamic driving force model [16] shown schematically in Figs. 3a and 3b. In the equilibrium state which corresponds to an as-deposited unstrained film, the free energy density of three grains shown in Fig. 3a can be considered to be virtually equal. Applied mechanical load will increase the free energy density by the amount of elastic strain energy density which can be different from grain to grain. The difference in elastic strain energy densities will result in the occurrence of a driving force, $\Delta P$, as shown in Fig. 3b. This driving force will promote the growth of grain 3 in the expense of grain 2 through the migration of the separating grain boundary.

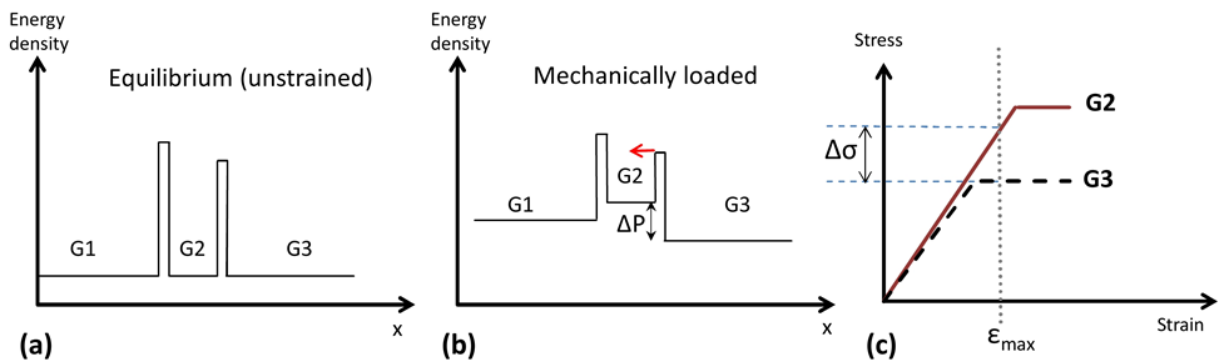

**Fig. 3.** Thermodynamic driving force leading to strain-induced grain coarsening. Diagrams (a) and (b) show schematically the free energy density of three grains (G1, G2, G3) in the unstrained and mechanically loaded state, respectively. The energy barriers between the grains represent the grain boundaries. In equilibrium the energy density within the grains is equal. When mechanical load is applied the free energy density increases by the amount of elastic strain energy density which can be different in each grain. This difference gives rise to the appearance of the driving force $\Delta P$ which results in the movement of the grain boundary between G2 and G3 as shown in (b) by the arrow. In (c) the schematic diagram of size-dependent yielding is presented. Larger grain G3 deforms plastically at lower stress than the smaller grain G2. The difference in the yield stress $\Delta \sigma$ leads to the difference in stored elastic energy density.

The difference in elastic strain energy density between the grains can appear due to several possible reasons. Since gold is an elastically anisotropic material, some grains will have a lower Young's modulus in the straining direction leading to a lower strain energy density at the same applied external strain [17,18]. Another possibility is that some grains will have a lower Taylor factor and plastic slip will occur at a lower applied stress. In both cases, the favored grains which coarsen during straining would have some specific orientations with respect to the loading axis which was not observed in the present study. Elastic stiffnesses and Taylor factors of the microstructure shown in Fig. 2a were also directly calculated and are depicted in Supplementary Material proving that the numbered grains do not exhibit specific values of these parameters. Another possible driving force can be induced by the differences in dislocation density [16,19]. However, in order to realize a sufficient driving force one should assume that the favored grains have an extremely low dislocation density while the other grains have a very high dislocation density [19]. There are no plausible reasons for such assumptions, especially taking into account that the generation and elimination of dislocations occurs permanently during cyclic mechanical loading.

We believe that the driving force responsible for the grain coarsening appears from the dependence of the yield stress on the grain size, generally known as Hall-Petch effect [20]. The schematic diagram illustrating the appearance of driving force due to size-dependent yielding is shown in Fig. 3c. Assuming perfect elastic-plastic behavior, the smaller grain 2 will have a higher stress state than the larger grain 3 which yields at a lower stress value. Consequently, the elastic strain energy density within grain 3 will be lower than within grain 2. Exact quantitative calculation of the stress difference is difficult due to the strain rate sensitivity and large spread of the experimentally measured yield stresses of UFG and NC gold [21-25]. For a rather rough estimate one can consider the yield stress of 200 MPa for the grain sizes above 400 nm [21,23] and 400 MPa for the grain sizes about 100 nm [21,24]. Such stress difference will give rise to a driving force of the order of $10^6$ J/m$^3$. This driving force is considered to be high enough to cause GB migration and grain coarsening. For comparison, the driving forces responsible for texture transition and recrystallization in thin films were estimated to be below $4 \times 10^5$ J/m$^3$ for copper films [18] and below $10^6$ J/m$^3$ for silver films [19]. Thus, the model presented in Fig. 3 fully explains the empirical observations, namely, isotropic preferable growth of initially larger grains independent of grain orientation or misorientation angle to the neighboring grains.

The exact atomistic mechanism of GB migration cannot be revealed directly in the frameworks of a thermodynamic model. In-situ high-resolution transmission electron microscopy studies suggest that GB migration occurs by propagation of atomic-size steps along the grain boundaries [13,26-29]. This mechanism was observed in different materials under different conditions, for example, in Au at room temperature without external strain [26], in Cu, Au and Al at high temperatures without external strain [27-29], and in Al at high temperature with applied strain [13]. The migration of GBs through the movement of atomic steps is consistent with the schematic energy diagram shown in Fig. 3b. The atoms at a grain boundary step edge are bonded to the crystal lattice more weakly than the atoms in a flat, step-free boundary. Consequently, the step edge atoms experience a lower energy barrier and have a greater chance to shuffle through the boundary [28].

Despite the internal consistency between the experimental observations and the model there is not enough experimental data to state that the size-dependent yielding is the *only* driving force which governs the grain coarsening. In order to estimate the generality of the model it must be clarified whether it is valid for different materials, microstructures, and fabrication methods. Further investigations are also needed to elucidate whether and how the dislocation activity is related to the grain coarsening.

In summary, detailed investigation of room temperature grain coarsening induced by cyclic strain in thin gold films is presented. It is shown that the favored grains which coarsen during mechanical loading have neither a specific orientation with respect to the straining direction nor a particular misorientation to the neighboring grains. The grain coarsening is also shown to be isotropic indicating that the grain boundaries migrate in all directions without correlation with the direction of applied strain. The single parameter which was found to be common for the favored grains is the initial grain size. Observed experimental data is explained within the frameworks of a thermodynamic driving force model where isotropic driving force appears due to the dependence of the yield stress of individual grains on the grain size. This model is consistent with the atomistic mechanism of grain boundary migration based on the motion of atomic-size steps along the boundary.

**Acknowledgements**

This work was supported by the Austrian Science Fund (FWF) through the projects P22648-N20 and P27432-N20. The authors would like to thank G. Dehm (MPIE Düsseldorf) for valuable discussions.

**References**

[1] K. Zhang, J.R. Weertman, J.A. Eastman, Appl. Phys. Lett. 87 (2005) 061921
[2] H. W. Höppel, Z. M. Zhou, H. Mughrabi, R.Z. Valiev, Phil. Mag. A. 81 (2002) 1781-1794
[3] L. Kunz, P. Lukáš, L. Pantelejev, O. Man, Procedia Eng. 10 (2011) 201–206
[4] R.A. Meirom, D. Hein Alsem, A. L. Romasco, T. Clark, R. G. Polcawich, J. S. Pulskamp, M. Dubey, R. O. Ritchie, C. L. Muhlstein, Acta Mater. 59, (2011) 1141-1149
[5] B.L. Boyce, H.A. Padilla II, Metall. Mater. Trans. 42 (2011) 1793-1804
[6] M. Jin, A. M. Minor, E. A. Stach, J. W. Morris Jr, Acta Mater. 52 (2004) 5381-5387
[7] D. S. Gianola, S. Van Petegem, M. Legros, S. Brandstetter, H. Van Swygenhoven, K.J. Hemker, Acta Mater. 54 (2006) 2253-2263
[8] T. J. Rupert, D. S. Gianola, Y. Gan, K. J. Hemker, Science 326 (2009) 1686-1690
[9] X. M. Luo, X. Zhu, G. P. Zhang, Nat. Commun. 5 (2014) 3021
[10] J. W. Cahn, Y. Mishin, A. Suzuki, Acta Mater. 54 (2006) 4953-4975
[11] T. Gorkaya, D. Molodov, G. Gottstein, Acta Mater. 57 (2009) 5396-5405
[12] E. R. Homer, S. M. Foiles, E. A. Holm, D. L. Olmsted, Acta Mater. 61 (2013) 1048-1060
[13] A. Rajabzadeh, M. Legros, N. Combe, F. Mompiou, D. A. Molodov, Phil. Mag. 93 (2013) 1299-1316
[14] F. Mompiou, D. Caillard, M. Legros, Acta Mater. 57 (2009) 2198-2209
[15] P. Wang, X. Yang, D. Pengm, Comp. Mater. Sci. 112 (2016) 289-296
[16] G. Gottstein, L. S. Shvindlerman, Grain Boundary Migration in Metals: Thermodynamics, Kinetics, Applications 2nd edn CRC Press (2010)
[17] R. Carel, C. V. Thompson, H.J. Frost, Acta Mater. 44 (1996) 2479-2494
[18] P. Sonnweber-Ribic, P. A. Gruber, G. Dehm, H. P. Strunk, E. Arzt, Acta Mater. 60 (2012) 2397-2406
[19] E. A. Ellis, M. Chmielusa, M.-T. Lin, H. Joress, K. Visser, A. Woll, R. P. Vinci, W. L. Brown, S. P. Baker, Acta Mater. 105 (2016) 495-504
[20] C. S. Pande, K. P. Cooper, Prog. Mater. Sci. 54 (2009) 689-706
[21] D. Son, J. H. Jeong, D. Kwon, Thin Solid Films 437 (2003) 182-187

[22] R. D. Emery, G. L. Povirk, Acta Mater. 51 (2003) 2079-2087
[23] Z. Gan, Y. He, D. Liu, B. Zhang, L. Shen, Scr. Mater. 87 (2014) 41-44
[24] G. D. Sim, J. J. Vlassak, Scr. Mater. 75 (2014) 34-37
[25] K. Jonnalagadda, N. Karanigaokar, I. Chasiotis, J. Chee, D. Peroulis, Acta Mater. 58 (2010) 4674-4684
[26] K. L. Merkle, L. J. Thompson, Mater. Lett. 78 (2001)188-193
[27] C. M. F. Rae, D. A. Smith, Phil. Mag. A, 47 (1980) 477-492
[28] S. E. Babcock, R. W. Baluffi, Acta Metall. 37 (1989) 2367-2376
[29] K. L. Merkle, L. J. Thompson, F. Phillipp, Interface Sci. 12 (2004) 277-292